\documentclass[12pt]{article}
\usepackage[top = 2 cm, bottom = 2 cm, left = 2.5 cm, right = 2 cm]{geometry}
\usepackage{authblk, cite, color, amssymb, amsmath, graphicx}
\usepackage[hypertex, colorlinks = true, linkcolor = blue, citecolor = red]{hyperref}
 \usepackage{booktabs}
 \usepackage[dvipsnames]{xcolor}

\begin{document}

\title{Gravitational Waves from Mirror World}

\author{{\bf Revaz Beradze}$^{1,2}$ and {\bf Merab Gogberashvili$^{1,3}$}}
\affil{\small $^1$ Javakhishvili Tbilisi State University, 3 Chavchavadze Avenue, Tbilisi 0179, Georgia \authorcr
$^2$ Universit`a dell’Aquila, Via Vetoio, 67100 Coppito, L’Aquila, Italy \authorcr
$^3$ Andronikashvili Institute of Physics, 6 Tamarashvili Street, Tbilisi 0177, Georgia}
\maketitle

\begin{abstract}

In this paper we consider the properties of the 10 confirmed by the LIGO Collaboration gravitational wave signals from the black hole mergers. We want to explain non-observation of electromagnetic counterpart and higher then expected merging rates of these events, assuming the existence of their sources in the hidden mirror universe. Mirror matter, which interacts with our world only through gravity, is a candidate of dark matter and its density can exceed ordinary matter density five times. Since mirror world is considered to be colder, star formation there started earlier and mirror black holes had more time to pick up the mass and to create more binary systems within the LIGO reachable zone. In total, we estimate factor of 15 amplification of black holes merging rate in mirror world with respect to our world, which is consistent with the LIGO observations.

\vskip 2mm


\end{abstract}


\section{Introduction}

One of the most important scientific achievements of the 21st century is the rise of the Gravitational Waves (GW) and multi-messenger astronomy. After the upgrades to the Advanced LIGO (Laser Interferometer Gravitational-Wave Observatory), the detector reached the sensitivity, which appeared enough to directly detect GWs for the first time. During the first observing run (O1), with runtime of 4.26 months (12~September 2015--19 January 2016), advanced LIGO observed two confirmed signals form merging Binary Black Holes (BBH) GW150914~\cite{signal1}, GW151226~\cite{signal2} and one candidate event LVT151012~\cite{candidate}. During the second observing run (O2) (30~November  2016--25 August 2017), with advanced VIRGO joining on 1 August, it has detected three more events GW170104~\cite{signal3}, GW170608~\cite{signal}, GW170814~\cite{signal4} from BBH system and one signal from merging neutron stars GW170817~\cite{signaln}. Later on, novel methods for reanalyzing of first two runs, revealed four more GW signals from BBHs (GW170729, GW170809, GW170818 and GW170823~\cite{signalnew}). In addition, the candidate event LVT151012~\cite{candidate} was promoted as a confident signal GW151012~\cite{signalnew}. In total, during these two runs LIGO observed 10~GWs from BBH mergers and one signal from binary neutron stars. It is important that the neutron star event and two BBH signals (GW170814 and GW170818) were triple-coincidence events, observed by two LIGO observatories together with VIRGO detector.

Table \ref{table} shows masses of the BHs and corresponding redshifts of 10 detected GWs signals obtained after reanalysis of data in~\cite{signalnew}. BBH merging rate, depended on these 10 signals and total runtime of O1 and O2, was estimated to be
\begin{equation} \label{LIGO}
\mathcal{R}_{\rm LIGO} = 9.7-101~ \rm Gpc^{-3} \rm yr^{-1} ~,
\end{equation}
with 90\% confidence~\cite{signalnew}. From these events GW170729 had the highest total mass $85.1^{+15.6}_{-10.9}$ and was located further, with luminosity distance $2750^{+1350}_{-1320}$~Mpc. The closest BBH event was GW170608, with luminosity distance $320^{+120}_{-110}$, having the smallest total mass.

Neutron star merging event was accompanied by the electromagnetic radiation, $\gamma$-ray burst detected by Fermi~\cite{GRB}. While, for now, it was unable to witness electromagnetic counterpart of the GWs from BH mergers~\cite{GRBBH, GRBBH2}, which leads to the idea that this were isolated BBHs---not surrounded by baryonic matter, which seems unnatural. That is why we want propose that the gravitational radiation detected by LIGO may have come from the hidden sector of our Universe, from the so-called Mirror World. Assuming that all Standard Model particles have their Mirror partners, the left-right symmetry of the nature can be restored~\cite{Blinnikov:1982}. However, Ordinary and Mirror particles interact mostly due to gravity and on the astrophysical scales, Mirror matter is supposed to exist in the form of stars and clusters like the Ordinary matter~\cite{Blinnikov:1983}.
\begin{table}
\caption {Masses of merging black holes (BHs) (in the units of solar mass) and corresponding redshifts.}
\centering
\begin{tabular}{ccccc}
\toprule
\textbf{\#}	& \textbf{Gravitational Wave} & \boldmath{$M_{1}~(M_{\odot})$} & \boldmath{$M_{2}~(M_{\odot})$} & \textbf{Redshift}\\
\midrule
1 & GW150914 & $35.6^{+4.8}_{-3.0}$ & $30.6^{+3.0}_{-4.4}$ & $0.09^{+0.03}_{-0.03}$ \\[0.1cm]
2 & GW151012 & $23.3^{+14.0}_{-5.5}$ & $13.6^{+4.1}_{-4.8}$ & $0.21^{+0.09}_{-0.09}$ \\[0.1cm]
3 & GW151226 & $13.7^{+8.8}_{-3.2}$ & $7.7^{+2.2}_{-2.6}$ & $0.09^{+0.04}_{-0.04}$ \\[0.1cm]
4 & GW170104 & $31.0^{+7.2}_{-5.6}$ & $20.1^{+4.9}_{-4.5}$ & $0.19^{+0.07}_{-0.08}$ \\[0.1cm]
5 & GW170608 & $10.9^{+5.3}_{-1.7}$ & $7.6^{+1.3}_{-2.1}$ &  $0.07^{+0.02}_{-0.02}$ \\[0.1cm]
6 & GW170729 & $50.6^{+16.6}_{-10.2}$ & $34.3^{+9.1}_{-10.1}$ & $0.48^{+0.19}_{-0.20}$ \\[0.1cm]
7 & GW170809 & $35.2^{+8.3}_{-6.0}$ & $23.8^{+5.2}_{-5.1}$ & $0.20^{+0.05}_{-0.07}$ \\[0.1cm]
8 & GW170814 & $30.7^{+5.7}_{-3.0}$ & $25.3^{+2.9}_{-4.1}$ & $0.12^{+0.03}_{-0.04}$ \\[0.1cm]
9 & GW170818 & $35.5^{+7.5}_{-4.7}$ & $26.8^{+4.3}_{-5.2}$ & $0.20^{+0.07}_{-0.07}$ \\[0.1cm]
10 & GW170823 & $39.6^{+10.0}_{-6.6}$ & $29.4^{+6.3}_{-7.1}$ & $0.34^{+0.13}_{-0.14}$\\[0.1cm]
\bottomrule
\end{tabular} \label{table}
\end{table}


\section{Binary Black Holes}

LIGO discoveries of GWs proves the existence of ``heavy'' BHs $(\gtrsim$25 $M_{\odot})$, confirms that they can form binary systems and merge within the age of the universe~\cite{abbott}. However, the mechanism of these processes is not fully clear.

The most common way for creating a BH is gravitational collapse of a heavy star. The final step of evolution of the stars is still speculative~\cite{endlife}, but according to the most common description, by the time the star runs out of fuel, if its core mass remains heavy enough, it explodes, creating a supernova and leaving a BH as a stellar remnant.
However, the mass of the remnant BH is not in one-to-one match with the mass of the progenitor star. In order to create a BH with such a ``heavy'' mass, the progenitor star should have some special parameters. Stellar wind is the main reason by which a star loses its mass and its strength is found to be dependent on the metallicity of the star~\cite{abbott}. Low metallicity reduces opacity, radiation transport becomes easier and decreases wind strength. So only the stars with a certain amount of metal content (below $ \simeq 1/2~Z_{\odot} $) are capable of creating a ``heavy'' BH~\cite{abbott}.

In~\cite{ligonew}, astrophysical origins of BBHs (Binary Black Holes), that generated GWs detected by LIGO, are reviewed. Several channels for forming the BBHs are discussed. BBH can be created through common-envelope~\cite{isolated} or via chemically homogeneous evolution~\cite{chemically} from isolated massive binaries in the galactic field; BBH might be formed also in a dense stellar cluster by some dynamical processes~\cite{dynamical}; Finally, BBHs can have a primordial origin~\cite{dominant, sasaki2016}.


\subsection{Primordial Black Holes}

Some authors suggested explanations for the sources of GW signals by so-called Primordial Black Holes (PBH)~\cite{dominant, sasaki2016}. PBHs are BHs that could have been formed in the early universe, when no astrophysical objects existed yet. The most popular mechanism for PBH formation is the direct gravitational collapse of primordial density inhomogeneities. As in the very early times, the universe was radiation dominated and ordinary matter was not yet formed, we can think of PBH as the direct collapse of Dark Matter (DM) density fluctuations, and it is common to define a fraction of PBHs in~DM
\begin{equation} \label{f}
f_{\rm PBH} = \frac {\Omega_{\rm PBH}}{\Omega_{\rm DM}}~,
\end{equation}
where $\Omega_{\rm PBH}$ and  $\Omega_{\rm DM}$ are PBH and DM density parameters, respectively.

Existing constraints on $f_{\rm PBH}$ in different mass ranges from various experiments are reviewed in~\cite{review}. Relevant for LIGO (see Table \ref{table}) mass interval is
\begin{equation} \label{M-LIGO}
1~M_{\odot} < M < 50 ~M_{\odot}~.
\end{equation}

\noindent However, microlensing surveys say that PBHs in the mass range
\begin{equation}
10^{-7} ~M_{\odot} < M < 30~ M_{\odot}
\end{equation}
cannot fill dominant parts of $\Omega_{\rm DM}$~\cite{microl1, microl2}; higher masses PBHs ($43M_{\odot}$$\lesssim$) are excluded by wide binaries~\cite{wideb} and the mass range 1--100~$ M_{\odot}$ is constrained by the non-detection of CMB spectral distortion~\cite{CMB}. This suggests that upper bound of $f_{\rm PBH}$ is order of $10^{-4}$--$10^{-3}$ in the mass range (\ref{M-LIGO}).

Several models can be responsible for the explanation of LIGO signals using the PBHs. These~models differ with the value of $f_{\rm PBH}$ and use distinct mechanism for binary system formation:
\begin{enumerate}
\item When two PBHs accidentally pass each other with sufficiently small impact parameter, they can form BBHs due to energy loss by gravitational radiation~\cite{dominant}. In this scenario, in order to explain the event rate estimated by LIGO, the fraction of PBHs in DM (\ref{f}) is required to be the order of unity. This is in contradiction with the CMB (Cosmic Microwave Background) anisotropies, but in~\cite{dominant}, it is assumed that constraints from CMB require modeling of several complex physical processes and, therefore, could have a significant uncertainty.

\item A different mechanism for estimating a PBH merging rate was suggested in~\cite{sasaki1997}. Cosmic expansion pulls PBHs away from each other, while gravitation tries to keep them together. If~gravitational energy between two PBHs exceeds expansion energy, they start to free-fall on one another. However,~neighboring PBH can exert torque on their system, avoiding their head-on collision and forming an eccentric binary in this way. In~\cite{sasaki1997} it was assumed that PBHs are massive stellar halo objects (MACHO) with monochromatic mass function equal to $0.5~M_{\odot}$, PBHs are initially randomly distributed in space and $f_{\rm PBH} \approx 1$. After the LIGO discovery, this theory was rewritten for PBHs with mass $30~M_{\odot}$~\cite{sasaki2016} and (\ref{f}) was treated as a free parameter. It was derived that, in order to get a merging rate compatible with LIGO's estimates, $f_{\rm PBH}$ is required to be of order $10^{-4}$. Intriguingly, it appears close to the PBH abundance estimated from the lack of CMB spectral distortion. More~GWs data is needed to test this model.
\end{enumerate}


\subsection{Astrophysical Black Holes}

Currently, most models are concerned with astrophysical origin of LIGO's BBHs~\cite{ligonew}. These models estimate BBH merging rate as a function of the efficiency of BBH merging $\epsilon$, the distribution of times elapsed between creating and merging of a binary system $P$ and BH's number density $N_{\rm BH}$~\cite{bbhnumber},
\begin{equation} \label{R}
\mathcal{R} = \frac{1}{2}~\epsilon P(\tau) N_{\rm BH} ~.
\end{equation}

\noindent Dimensionless coefficient
\begin{equation}
\epsilon \equiv f_{\rm bin} \times f_{\rm m_1/m_2} \times f_{\rm surv} \times f_{\rm t} <1
\end{equation}
defines the efficiency of BBH merging~\cite{bbhnumber}. Current models predict that half of the stars are in binaries $f_{\rm bin} \sim 0.5$~\cite{Sana}, $f_{\rm m_1/m_2} \sim 0.1$ is the fraction of binary systems of stars, which have the mass ratio near unity $m_1/m_2 \sim 1$~\cite{Sana}, which corresponds to the LIGO data in Table \ref{table}, and  $f_{\rm surv} \sim 0.1$ is fraction of massive stars that survive as BH pairs after stellar evolution. Finally, $f_{\rm t}<1$ is a fraction of BBHs with orbital configuration that makes them available to merge before the present day. As we see, $\epsilon$ depends on many factors and can vary significantly in the interval
\begin{equation}
\epsilon \simeq 0.01-0.001~.
\end{equation}

Delay time $P(\tau)$ is also very speculative as it depends on the masses, metallicities, orbital configurations of the binary system of progenitor stars and it can even exceed the Hubble time. BH number density can be written as~\cite{bbhnumber}
\begin{equation} \label{NBH}
N_{\rm BH} = {\rm SFR(z)} \int \phi(m)~N(m)\int f(Z,m) \int \xi(M)~dM~dZ~dm~ .
\end{equation}

\noindent Here $\xi(M)$ is a stellar initial mass function (which usually is integrated in the interval $5~ M_{\odot} < M <150 ~M_{\odot}$ to match the LIGO data), $f(Z,m)$ is a metallicity distribution function of the galaxy of the mass $m$ (usually the metallicity range $0.0002 < Z < 0.02$ is considered), $N(m)$ is total number of stars in the galaxy of mass $m$, which is normalized as
\begin{equation} \label{N}
N(m)=\frac{m}{\int M~\xi(M)~dM}~ ,
\end{equation}
 $\phi(m)$ is a galactic stellar mass function and SFR(z) is a star formation rate, which is typically adopted from the best-fit-function of experimental data~\cite{SFR},
\begin{equation} \label{SFR}
{\rm SFR(z)} = 0.015 \frac{(1+z)^{2.7}}{1+[(1+z)/2.9]^{5.6}} ~ \rm M_{\odot} ~ Mpc^{-3} ~ yr^{-1}~.
\end{equation}

\noindent This function peaks at $z \sim 2$ corresponding to the luminosity distance $\sim$15~Mpc and lookback time~$\sim$10~Gyr.

In the scenario for isolated binaries formed throughout common envelope evolution~\cite{isolated}, simulations were carried out for different values of metallicities. Derived local BBH merging rate spans from $\sim$10$^{-1}$ to $7 \times 10^3 ~ \rm Gpc^{-3}~yr^{-1}$. Such a high uncertainty comes from tight dependence on metallicity distribution function of progenitor stars. In order to fall in merging rate estimated by LIGO, lower values of metallicities are favored.

Simulations for chemically homogeneous stellar binaries~\cite{chemically} suggest $\mathcal{R} \approx 10~ \rm Gpc^{-3}~yr^{-1}$. They~find that typical time delay between formation and merger $P(\tau)$ ranges from 4 to 11 Gyr and mergers beyond $z \gtrsim 1.6$ did not take place as the Universe was too young. They conclude that over cosmic time, merger rate rises as mergers with longer delay times start to contribute, but in the present age Universe it start to fall, as low-metallicity SFR decreases, leading to a peak of $\sim$20~$\rm Gpc^{-3}~yr^{-1}$ at $z \lesssim 0.5$.

In~\cite{dynamical} thousands of dense star cluster models with different initial conditions were simulated and coalescing BBHs that escaped or merged inside the clusters were studied. The local merger rate density of BBHs originated from globular clusters is obtained to be $5.4~\rm Gpc^{-3}~yr^{-1}$.

All these models~\cite{isolated, chemically, dynamical} estimate a theoretical BBH merger rate
\begin{equation} \label{theor}
 \mathcal{R}_{\rm theor} \sim 5-10 ~ {\rm Gpc^{-3}~yr^{-1}}~,
\end{equation}
which is near the LIGO's lower bound (\ref{LIGO}).


\section{Gravitational Waves from the Mirror World}

Existing scenarios for explaining the BBH merging rate lack some confidence. Primordial BHs are ruled out as LIGO's GW sources by CMB and microlensing experiments. Astrophysical binary BH models predict the merging rate (\ref{R}) near the LIGO's lower bound (\ref{theor}), but they require low metallicities of the progenitor stars and certain delay times for binary system merging and uncertainties in $P$ and $\epsilon$ can vary significantly~\cite{isolated, chemically, dynamical}.

In this paper we want to suggest a new explanation for LIGO BHs events using Mirror World (M-World) scenario~\cite{Khlopov:1989fj}: the BBH systems that produced GWs could have existed in Mirror, or Parallel World, which interacts with our world only through gravity. Then GWs from binary BHs had no electromagnetic counterparts in our world, because mirror photons cannot interact with ordinary matter. M-World is a possible candidate of DM, and as DM is up to five times more that the ordinary matter in the Universe, it can increase BBH merger rate naturally. Before estimating M-World BBH merger rate, let us briefly describe the basic concepts of the M-World and its cosmological implications; for details see the review~\cite{mirror}.


\subsection{Mirror World}

M-World initially was introduced to restore a left-right symmetry of the nature~\cite{Khlopov:1989fj, mirror}. In~Standard Model of particle physics, only left-handed particles participate in weak interactions and currents have (V-A) type. In M-World, chiralities are opposite---right-handed particles interacting through weak force, with right (V+A) currents. Then it is assumed that each O-World particle has its Mirror partner, which is invisible for observable from our world and visa-versa. Such a theory can naturally emerge in the context of the heterotic string theory~\cite{Kolb:1985bf, Khl-Shi}, based, for example, on the $E_8 \times E'_8$  gauge group. In such group $E_8 \leftrightarrow E'_8$ symmetry can originate two forms of matter: ordinary and shadow, with the interactions described by the gauge groups $E_8$ and $E'_8$, respectively. The particular case of the shadow world can be the mirror world. One can consider a theory with two identical gauge groups $G \times G'$ and with identical particle concept. So if $G$ is symmetry group of O-World physics, e.g., $SU(3) \times SU(2) \times U(1)$ in Standard Model, the symmetry group $G' = SU(3)' \times SU(2)' \times U(1)'$ corresponds to the M-World. Hereafter we denote M-World parameters with primed qualities. Mirror particles are singlets of O-World and vice versa, i.e., ordinary particles are M-World singlets. The~only~possibility for the interaction between these two worlds is gravity and maybe some other unknown weak forces. In addition, kinetic mixing of ordinary and mirror photons, and oscillation of neutral particles (for example neutron) between these sectors are possible~\cite{mirror}.

We can imagine the M-World scenario as a five-dimensional theory, with parallel 3D-branes located in two fixed points; Ordinary matter being localized on the left-brane and Mirror matter localized on the right-brane, while gravity can freely pass between these two branes.

If mirror sector exists, it was also created by the Big Bang, along with the ordinary matter. However,~cosmological abundance of ordinary and mirror particle and their cosmological evolution cannot be identical. Big Bang Nucleosynthesis (BBN) bounds effective number of extra light neutrinos, $\Delta N_{\nu}<1$~\cite{deltaN}, and mirror particles would also contribute in the Hubble expansion rate equivalent to $\Delta N_{\nu} \simeq 6.14$~\cite{berez.vil}. In order to reduce M-particle density in the early universe and make their contribution in the Hubble expansion negligible, M-World should have had a lower reheating temperature than~O-World,
\begin{equation}
T'_R<T_R~,
\end{equation}
 which can be achieved in some inflationary models~\cite{reheating}. If at early times temperatures of two worlds are different and they interact very weakly (through gravity), they cannot come into thermal equilibrium. Therefore, these worlds will evolve independently during the cosmological evolution and at later stages maintain nearly constant temperature ratio, denoted by
\begin{equation} \label{tratio}
x \equiv \frac {T'}{T}~.
\end{equation}

\noindent BBN constraint $\Delta N_{\nu}<1$, sets upper bound for the ratio (\ref{tratio})
\begin{equation}
x < 0.64~.
\end{equation}

\noindent This condition, in the context of the GUT or electroweak baryogenesis scenarios, implies that baryon asymmetry $\eta_b'=n_b'/n_{\gamma}'$ in M-World is greater than in O-World $\eta_b=n_b/n_{\gamma}$, where $n_b$, $n_{\gamma}$, $n_b'$ and $n_{\gamma}'$ are the number densities of baryons and photons in O- and M-Worlds, respectively~\cite{berez.vil}. However, $\eta'_b / \eta_b  \geq 1$ does not directly mean that the ratio $n'_b/n_b \geq 1$, but in certain leptogenesis scenarios suggested in~\cite{leptogenesis}, the value
\begin{equation}
1 \leq \frac {n'_b}{n_b}  \lesssim 10
\end{equation}
can be achieved. If one considers mirror baryon matter as DM candidate, this ratio explains near coincidence between visible matter density $\Omega_b$ and dark (M-baryon) matter density $\Omega'_b$ without fine tuning. In general, one can assume that non-relativistic matter content of the universe consists of ordinary (visible) baryons, mirror baryons and other DM candidate (e.g., Cold Dark Matter---CDM)
\begin{equation}
\Omega_m = \Omega_b + \Omega'_b + \Omega_{\rm CDM}~.
\end{equation}

\noindent Current observations suggest that the universe is almost flat, with energy density very close to critical
\begin{equation}
\Omega_m + \Omega_r + \Omega_{\Lambda} \approx 1~.
\end{equation}

\noindent Here cosmological term $\Omega_{\Lambda} \simeq 0.73$,  $\Omega_m \approx 0.27$ (with $\Omega_b \approx 0.044$), while radiation, $\Omega_r$, gives negligible contribution. Leptogenesis mechanism~\cite{leptogenesis} can imply
\begin{equation} \label{mratio}
\frac {\Omega'_b}{\Omega_b} \approx 5 ~,
\end{equation}
which means that all DM can be explained by the mirror baryons, leaving no place for other DM candidates ($\Omega_{\rm CDM}=0$)~\cite{cosmological}.

The important features of structure formation are related to the recombination and matter-radiation decoupling (MRD) epoch. MRD in ordinary universe happens at the temperature $T_{\rm dec} \simeq 0.26 ~ \rm eV$, in the matter domination period, which corresponds to the redshift
\begin{equation}
1+z_{\rm dec} \simeq 1100
\end{equation}

\noindent However, MRD in M-Universe occurs earlier~\cite{cosmological}
\begin{equation}
1+z'_{\rm dec} \simeq x^{-1} (1+z_{\rm dec}) \simeq 2500
\end{equation}
and can take place even in the radiation domination era.


\subsection{BBH Merger Rate in M-World}

As we have seen, the mirror world evolves alongside our world, with the difference that it has lower temperature and so all the epochs and processes occur earlier. This means that the star formation rate (\ref{SFR}) will peak earlier at $z \sim$  4--6 (look back time in our world $\sim$12~Gyr) depending on the temperature ratio (\ref{tratio}). This implies that the SFR in mirror world is maximal at the luminosity distance $\sim$35--55~Mpc and so mirror BHs have more time to pick up mass and to create binaries in the area covered by the LIGO observations. Compering the difference in the luminosity distances to the maximums of the function (\ref{SFR}) for two worlds, we estimate factor of 3 amplification of star formation rate in mirror world relative to our world

\begin{equation}
{\rm SFR}'(z) \sim 3 \times {\rm SFR}(z)~.
\end{equation}

\noindent The additional argument of having more BH formed in mirror sector is that heavy mirror stars evolve much faster than ordinary ones of the same mass~\cite{Berezhiani:2005vv}, and their way to BH should be faster. In addition, after explosion of mirror supernovas, the ejected materials will be reprocessed again and can form new heavy stars.

Besides that, as we mentioned above, the mirror matter density can be five times greater than the ordinary matter density (\ref{mratio}). From (\ref{N}), this suggests five times bigger star abundance in mirror galaxy
\begin{equation}
N'(m) \sim 5 \times N(m) ~,
\end{equation}
and so, in total we derive factor of $\sim 15$ bigger BH number density (\ref{NBH}) in mirror world relative to our~world
\begin{equation}
N'_{\rm BH} \sim 15 \times N_{\rm BH}~.
\end{equation}

\noindent As a consequence we get factor of $\sim 15$ amplification of BBH merger rate (\ref{R}) in the mirror world compared to the ordinary world.  Adopting typical theoretical values from different models (\ref{theor}), we~estimate
\begin{equation}
\mathcal{R}_{\rm mirror}  \sim 15 \times \mathcal{R}_{\rm theor} \sim 75-150 ~ {\rm Gpc^{-3}~yr^{-1}}.
\end{equation}

\noindent So our analysis gives BBH merger rate density in the upper interval of LIGO's measurements (\ref{LIGO}).


\section{Conclusions}

To conclude, inspired by the fact that the 10 currently confirmed GW signals from BBH mergers during relatively small observational period had no counterpart electromagnetic radiation, in this paper we explore the idea that the sources of these events existed in the hidden mirror universe, which interacts with our world only through gravity. Mirror matter is a candidate of dark matter and its density can exceed ordinary matter density five times. Besides that, since mirror world is considered to be colder, star formation there started earlier and its rate peaks are at greater $z$. In total, we estimated factor of 15 amplification of merger rate in mirror world with respect to our world. Adopting a common approach, we derived the BBH merging rate 75--150~ $\rm Gpc^{-3}~yr^{-1}$, which is in good agreement with the LIGO observations.

\vspace{10pt}
\noindent
{\bf Acknowledgements:} This work was supported by Shota Rustaveli National Science Foundation of Georgia (SRNSFG) [DI-18-335/New Theoretical Models for Dark Matter Exploration].



\begin{thebibliography}{99}

\bibitem{signal1}
Abbott, B.P.; {et al.} [LIGO Scientific and Virgo Collaborations].
Observation of Gravitational Waves from a Binary Black Hole Merger.
\emph{Phys. Rev. Lett.} \textbf{2016}, {\em 116}, 061102, arXiv: 1602.03837 [gr-qc]. 
doi:10.1103/PhysRevLett.116.061102

\bibitem{signal2}
Abbott, B.P.; {et al.}  [LIGO Scientific and Virgo Collaborations].
GW151226: Observation of Gravitational Waves from a 22-Solar-Mass Binary Black Hole Coalescence.
\emph{Phys.\ Rev.\ Lett.}  \textbf{2016}, {\em 116}, 241103, arXiv: 1606.04855 [gr-qc]. 
doi:10.1103/PhysRevLett.116.241103

\bibitem{candidate}
Abbott, B.P.; {et al.} [LIGO Scientific and Virgo Collaborations].
Binary Black Hole Mergers in the first Advanced LIGO Observing Run.
\emph{Phys.\ Rev.\ X} \textbf{2016}, {\em 6}, 041015;
Erratum in \emph{Phys.\ Rev.\ X} \textbf{2018}, {\em 8}, 03990, arXiv: 1606.04856 [gr-qc].
doi:10.1103/PhysRevX.6.041015, 10.1103/PhysRevX.8.039903

\bibitem{signal3}
Abbott, B.P.; {et al.} [LIGO Scientific and VIRGO Collaborations].
GW170104: Observation of a 50-Solar-Mass Binary Black Hole Coalescence at Redshift 0.2.
\emph{Phys.\ Rev.\ Lett.} \textbf{2017}, {\em 118},  221101;
Erratum in \emph{Phys.\ Rev.\ Lett.} \textbf{2018}, {\em 121}, 129901, arXiv: 1706.01812 [gr-qc].
doi:10.1103/PhysRevLett.118.221101, 10.1103/PhysRevLett.121.129901

\bibitem{signal}
Abbott, B.P.; {et al.} [LIGO Scientific and Virgo Collaborations].
GW170608: Observation of a 19-solar-mass Binary Black Hole Coalescence.
\emph{Astrophys.\ J.} \textbf{2017}, {\em 851}, L35, arXiv: 1711.05578 [astro-ph.HE].
doi:10.3847/2041-8213/aa9f0c

\bibitem{signal4}
Abbott, B.P.; {et al.} [LIGO Scientific and Virgo Collaborations].
GW170814: A Three Detector Observation of Gravitational Waves from a Binary Black Hole Coalescence.
\emph{Phys.\ Rev.\ Lett.} \textbf{2017}, {\em 119}, 141101, arXiv: 1709.09660 [gr-qc].
doi:10.1103/PhysRevLett.119.141101

\bibitem{signaln}
Abbott, B.P.; {et al.} [LIGO Scientific and Virgo Collaborations].
GW170817: Observation of Gravitational Waves from a Binary Neutron Star Inspiral.
\emph{Phys.\ Rev.\ Lett.}  \textbf{2017}, {\em 119}, 161101, arXiv: 1710.05832 [gr-qc].
doi:10.1103/PhysRevLett.119.161101

\bibitem{signalnew}
Abbott. B.P.; et al.  [LIGO Scientific and Virgo Collaborations].
GWTC-1: A Gravitational-Wave Transient Catalog of Compact Binary Mergers Observed by LIGO and Virgo during the First and Second Observing Runs. \emph{High Energy Astrophys. Phenom.} \textbf{2018},  arXiv:1811.12907.

\bibitem{GRB}
Goldstein, A.; Veres, P.; Burns, E.; Briggs, M.S.; Hamburg, R.; Kocevski, D.; Wilson-Hodge, C.A.; Preece, R.D.; Poolakkil, S.; Roberts, O.J.; et al.
An Ordinary Short Gamma-Ray Burst with Extraordinary Implications: Fermi-GBM Detection of GRB 170817A.
\emph{Astrophys.\ J.} \textbf{2017}, {\em 848}, L14, arXiv: 1710.05446 [astro-ph.HE].  
doi:10.3847/2041-8213/aa8f41

\bibitem{GRBBH}
Burns, E.; {et al.} [LIGO Scientific and Virgo Collaborations and Fermi Gamma-ray Burst Monitor Team].
A Fermi Gamma-ray Burst Monitor Search for Electromagnetic Signals Coincident with Gravitational- Wave Candidates in Advanced LIGO's First Observing Run. \emph{High Energy Astrophys. Phenom.} \textbf{2018},
arXiv:1810.02764.

\bibitem{GRBBH2}
Racusin,   J.L. {et al.} [Fermi-LAT Collaboration].
Searching the Gamma-ray sky for Counterparts to Gravitational Wave Sources: Fermi Gamma-ray Burst Monitor and Large Area Telescope Observations of LVT151012 and GW151226.
\emph{Astrophys.\ J.}  \textbf{2017}, {\em 835}, 82, arXiv: 1606.04901 [astro-ph.HE]. 
doi:10.3847/1538-4357/835/1/82

\bibitem{Blinnikov:1982}
Blinnikov, S.I.;  Khlopov,  M.Y.
On Possible Effects Of 'Mirror' Particles.
\emph{Sov.\ J.\ Nucl.\ Phys.} \textbf{1982}, {\em 36}, 472; translated from
\emph{Yad.\ Fiz.}  \textbf{1982}, {\em 36}, 809.

\bibitem{Blinnikov:1983}
Blinnikov, S.I.;   Khlopov, M.
Possible Astronomical Effects of Mirror Particles.
\emph{Sov.\ Astron.}  \textbf{1983}, {\em 27}, 371; translated from
\emph{Astron.\ Zh.} \textbf{1983}, {\em 60}, 632.

\bibitem{abbott}
Abbott, B.P. {et al.} [LIGO Scientific and Virgo Collaborations].
Astrophysical Implications of the Binary Black-Hole Merger GW150914.
\emph{Astrophys.\ J.} \textbf{2016}, {\em 818}, L22, arXiv: 1602.03846 [astro-ph.HE].  
doi:10.3847/2041-8205/818/2/L22

\bibitem{endlife}
Heger, A.; Fryer, C.L.; Woosley,  S.E.; Langer, N.;   Hartmann, D.H.
How Massive Single Stars end Their Life.
\emph{Astrophys.\ J.} \textbf{2003}, {\em 591}, 288--300, arXiv: astro-ph/0212469.
doi:10.1086/375341

\bibitem{ligonew}
Abbott, B.P. {et al.} [LIGO Scientific and Virgo Collaborations].
Binary Black Hole Population Properties Inferred from the First and Second Observing Runs of Advanced LIGO and Advanced Virgo.  \emph{High Energy Astrophys. Phenom.} \textbf{2018},
arXiv:1811.12940.

\bibitem{isolated}
Giacobbo, N.;  Mapelli, M.
The Progenitors of Compact-object Binaries: Impact of Metallicity, Common Envelope and Natal Kicks.
\emph{Mon.\ Not.\ R.\ Astron.\ Soc.} \textbf{2018}, {\em 480}, 2011, arXiv: 1806.00001 [astro-ph.HE].

\bibitem{chemically}
Mandel, I.;   de Mink, S.E.
Merging Binary Black Holes Formed Through Chemically Homogeneous Evolution in Short-period Stellar Binaries.
\emph{Mon.\ Not.\ R.\ Astron.\ Soc.} \textbf{2016}, {\em 458}, 2634, arXiv: 1601.00007 [astro-ph.HE].
doi:10.1093/mnras/stw379

\bibitem{dynamical}
Askar, A.; Szkudlarek, M.; Gondek-Rosińska, D.; Giersz, M.;   Bulik, T.
MOCCA-SURVEY Database---I. Coalescing Binary Black Holes Originating From Globular Clusters.
\emph{Mon.\ Not.\ R.\ Astron.\ Soc.}  \textbf{2017}, {\em 464}, L36.arXiv: 1608.02520 [astro-ph.HE].  
doi:10.1093/mnrasl/slw177

\bibitem{dominant}
Bird,  S.; Cholis, I.; Mu\~{n}oz,  J.B.; Ali-Ha\"{\i}moud, Y.; Kamionkowski, M.; Kovetz, E.D.;  Raccanelli, A.;   Riess, A.G.
Did LIGO Detect Dark Matter?
\emph{Phys.\ Rev.\ Lett.} \textbf{2016}, {\em 116}, 201301, arXiv: 1603.00464 [astro-ph.CO].
doi:10.1103/PhysRevLett.116.201301

\bibitem{sasaki2016}
Sasaki, M.; Suyama, T.; Tanaka, T.;   Yokoyama, S.
Primordial Black Hole Scenario for the Gravitational-Wave Event GW150914.
\emph{Phys.\ Rev.\ Lett.} \textbf{2016}, {\em 117}, 061101;
Erratum in  \emph{Phys.\ Rev.\ Lett.} \textbf{2018}, {\em 121}, 059901, arXiv: 1603.08338 [astro-ph.CO]. 
doi:10.1103/PhysRevLett.121.059901, 10.1103/PhysRevLett.117.061101

\bibitem{review}
Sasaki, M.; Suyama, T.; Tanaka, T.;   Yokoyama, S.
Primordial Black Holes---Perspectives in Gravitational Wave Astronomy.
\emph{Class.\ Quant.\ Grav.}  \textbf{2018}, {\em 35}, 063001, arXiv: 1801.05235 [astro-ph.CO]. 
doi:10.1088/1361-6382/aaa7b4

\bibitem{microl1}
Allsman, R.A.;  {et al.} [Macho Collaboration].
MACHO Project Limits on Black Hole Dark Matter in the 1--30 Solar Mass Range.
\emph{Astrophys.\ J.} \textbf{2001}, {\em 550}, L169, arXiv: astro-ph/0011506. 
doi:10.1086/319636

\bibitem{microl2}
Tisserand, P.; {et al.} [EROS-2 Collaboration].
Limits on the Macho Content of the Galactic Halo from the EROS-2 Survey of the Magellanic Clouds.
\emph{Astron.\ Astrophys.} \textbf{2007}, {\em 469}, 387--404, arXiv: astro-ph/0607207. 
doi:10.1051/0004-6361:20066017

\bibitem{wideb}
Yoo,  J.; Chaname, J.;  Gould, A.
The end of the MACHO Era: Limits on Halo Dark Matter from Stellar Halo Wide Binaries.
\emph{Astrophys.\ J.} \textbf{2004}, {\em 601}, 311--318, arXiv: astro-ph/0307437. 
doi:10.1086/380562

\bibitem{CMB}
Ricotti, M.; Ostriker, J.P.;  Mack, K.J.
Effect of Primordial Black Holes on the Cosmic Microwave Background and Cosmological Parameter Estimates.
\emph{Astrophys.\ J.}  \textbf{2008}, {\em 680}, 829, arXiv: 0709.0524 [astro-ph]. 
doi:10.1086/587831

\bibitem{sasaki1997}
Nakamura, T.; Sasaki, M.; Tanaka, T.;  Thorne, K.S.
Gravitational Waves from Coalescing Black Hole MACHO Binaries.
\emph{Astrophys.\ J.} \textbf{1997}, {\em 487}, L139, arXiv: astro-ph/9708060. 
doi:10.1086/310886

\bibitem{bbhnumber}
Elbert, O.D.; Bullock, J.S.; Kaplinghat, M.
Counting Black Holes: The Cosmic Stellar Remnant Population and Implications for LIGO.
\emph{Mon.\ Not.\ Roy.\ Astron.\ Soc.}  \textbf{2018}, {\em 473}, 1186, arXiv: 1703.02551 [astro-ph.GA].

\bibitem{Sana}
Sana, H.;   de Mink, S.E.; de Koter, A.; Langer, N.; Evans, C.J.; Gieles, M.; Gosset, E.; Izzard, R.G.; \mbox{le Bouquin, J}.; Schneider, B.F.R.N.
Binary Interaction Dominates the Evolution of Massive Stars.
\emph{Science} \textbf{2012}, {\em 337}, 444, arXiv: 1207.6397 [astro-ph.SR].
doi:10.1126/science.1223344

\bibitem{SFR}
Madau, P.;  Dickinson, M.
Cosmic Star Formation History.
\emph{Ann.\ Rev.\ Astron.\ Astrophys.}  \textbf{2014}, {\em 52}, 415, arXiv: 1403.0007 [astro-ph.CO].
doi:10.1146/annurev-astro-081811-125615

\bibitem{Khlopov:1989fj}
Khlopov, M.Y.; Beskin, G.M.; Bochkarev, N.E.; Pustylnik, L.A.;  Pustylnik, S.A.
Observational Physics of Mirror World.
\emph{Sov.\ Astron.}  \textbf{1991}, {\em 35}, 21; translated from
\emph{Astron.\ Zh.}  \textbf{1991}, {\em 68}, 42.

\bibitem{mirror}
Berezhiani, Z.
Through the Looking-glass: Alice's Adventures in Mirror World.
In \emph{From Fields to Strings};  Shifman, M., et al., Eds.; World Scientific: Singapore, 2005; Volume 3, pp.~2147--2195, arXiv: hep-ph/0508233. 
doi:10.1142/9789812775344\_0055

\bibitem{Kolb:1985bf}
Kolb, E.W.; Seckel, D.;  Turner, M.S.
The Shadow World.
\emph{Nature} \textbf{1985}, {\em 314}, 415.
doi:10.1038/314415a0

\bibitem{Khl-Shi}
Khlopov, M.Y.; Shibaev, K.I.
New physics from superstring phenomenology.
\emph{Grav. \ Cosmol. \ Suppl.} 2002, {\em 8},~45.

\bibitem{deltaN}
Lisi, E.; Sarkar, S.;  Villante, F.L.
The Big Bang Nucleosynthesis Limit on N(Neutrino).
\emph{Phys.\ Rev.\ D} \textbf{1999}, {\em 59},~123520, arXiv: hep-ph/9901404. 
doi:10.1103/PhysRevD.59.123520

\bibitem{berez.vil}
Berezhiani, Z.; Comelli, D.;  Villante, F.L.
The Early Mirror Universe: Inflation, Baryogenesis, Nucleosynthesis and Dark Matter.
\emph{Phys.\ Lett.\ B} \textbf{2001}, {\em 503}, 362, arXiv: hep-ph/0008105.

\bibitem{reheating}
Berezhiani, Z.G.; Dolgov, A.D.;  Mohapatra, R.N.
Asymmetric Inflationary Reheating and the Nature of Mirror Universe.
\emph{Phys.\ Lett.\ B}  \textbf{1996}, {\em 375}, 26, arXiv: hep-ph/9511221. 
doi:10.1016/0370-2693(96)00219-5

\bibitem{leptogenesis}
Bento, L.; Berezhiani, Z.
Leptogenesis via Collisions: The Lepton Number Leaking to the Hidden Sector.
\emph{Phys.\ Rev.\ Lett.} \textbf{2001}, {\em 87}, 231304, arXiv: hep-ph/0107281. 
doi:10.1103/PhysRevLett.87.231304

\bibitem{cosmological}
Berezhiani, Z.
Mirror World and its Cosmological Consequences.
\emph{ Int.\ J.\ Mod.\ Phys. A}  \textbf{2004}, {\em 19}, 3775, arXiv: hep-ph/0312335. 
doi:10.1142/S0217751X04020075

\bibitem{Berezhiani:2005vv}
Berezhiani,  Z.; Cassisi, S.;  Ciarcelluti, P.;  Pietrinferni, A.
Evolutionary and Structural Properties of Mirror Star MACHOs.
\emph{Astropart.\ Phys.} \textbf{2006}, {\em 24}, 495, arXiv: astro-ph/0507153. 
doi:10.1016/j.astropartphys.2005.10.002

\end{thebibliography}
\end{document}